\documentstyle[12pt]{article}
\def\cc{\hbox{\rm C\hskip-4.8pt\vrule height6.1pt\hskip5.3pt}}

\def\zz{\hbox{\rm Z\hskip -4pt Z}}
\def\eins{\hbox{\rm 1\hskip -3pt I}}
\font\eightrm                = cmr8
\font\eightsl                = cmsl8
\font\eightsy                = cmsy8
\font\eightit                = cmti8

\font\eighti                 = cmmi8
\font\eightbf                = cmbx8
\def\petit{\def\rm{\fam0\eightrm}
\textfont0=\eightrm %\scriptfont0=\sixrm \scriptscriptfont0=\fiverm
 \textfont1=\eighti %\scriptfont1=\sixi \scriptscriptfont1=\fivei
 \textfont2=\eightsy %\scriptfont2=\sixsy \scriptscriptfont2=\fivesy
 \def\it{\fam\itfam\eightit}
 \textfont\itfam=\eightit
 \def\sl{\fam\slfam\eightsl}
 \textfont\slfam=\eightsl
 \def\bf{\fam\bffam\eightbf}
 \textfont\bffam=\eightbf %\scriptfont\bffam=\sixbf
 \normalbaselineskip=9pt
 \setbox\strutbox=\hbox{\vrule height7pt depth2pt width0pt}
 \normalbaselines\rm}
\newdimen\refindent
\def\begref{\vskip1cm\bgroup\petit
\setbox0=\hbox{[Bi,Sc,So]o }\refindent=\wd0
\let\sl=\rm\let\INS=N}
\def\ref#1{\filbreak\if N\INS\let\INS=Y\vbox{\noindent\tbfontt
Refererences\vskip1cm}\fi\hangindent\refindent
\hangafter=1\noindent\hbox to\refindent{#1\hfil}\ignorespaces}

\def\ep{{\cal E}}

\def\mm{{\cal M}}

\def\di{\,\hbox{\rm D}}

\def\pa{{\partial}}
\def\dit{\tilde{\,\hbox{\rm D}}}

\def\ot{\otimes}
\def\op{\oplus}
\def\hot{\hat\otimes}

\def\bb{\begin{eqnarray}}
\def\ee{\end{eqnarray}}
\def\eee{\nonumber\end{eqnarray}}

\begin{document}

\hsize 17truecm
\vsize 24truecm
\font\twelve=cmbx10 at 13pt
\font\eightrm=cmr8
\baselineskip 18pt

\begin{titlepage}

\centerline{\twelve CENTRE DE PHYSIQUE THEORIQUE}
\centerline{\twelve CNRS - Luminy, Case 907}
\centerline{\twelve 13288 Marseille Cedex}
\vskip 3truecm

\centerline{\twelve Unification of Gravity and
Yang-Mills-Higgs Gauge Theories}

\bigskip

\begin{center}
{\bf Thomas Ackermann
\footnote{\indent Ackerm@euler.math.univ-mannheim.de}
\\ \bf J\"urgen Tolksdorf*
\footnote{\indent Tolkdorf@cptsu4.univ-mrs.fr \\
\indent *\indent Supported by by the European Communities,
 Contract No. ERB 401GT 930224}} \\
\end{center}

\vskip 1truecm
\leftskip=1cm
\rightskip=1cm

\medskip

In this letter we show how the action functional of the standard
model and of gravity can be derived from a specific Dirac operator.
Far from being exotic this particular Dirac operator turns out to be
structurally determined by the Yukawa coupling term. The main feature
of our approach is that it naturally unifies the action of the standard
model with gravity.

%Our approach to
%the full standard model fits in a general scheme which is presented here
%in a ``model building kit'' like form. This is analogous to the
%``Connes-Lott building kit'' but without using the mathematical
%frame work of non-commutative geometry.
%The Higgs field has a simple geometrical interpretation: it defines together
%with the usual Yang-Mills potential a particular connection on the space
%of fermions. As a consequence the representation of the Higgs field is
%not arbritrary but must be contained in the fermionic representation
%of the gauge group.

\vskip 6truecm

\noindent PACS-92: 11.15 Gauge field theories\par
\noindent MSC-91: 81E13 Yang-Mills and other gauge theories

\vskip 1truecm

\noindent march 1995
\vskip 0.1truecm
\noindent CPT-95/P.3180\\
\noindent Mannheimer Manuskripte 191\\
arch-ive/9503180

%\vskip 3truecm

\end{titlepage}

\section{Introduction}

\smallskip

Usually, the bosonic part of the standard model of elementary particles
is based on the Yang-Mills-Higgs action which is interpreted as a
specific functional ${\cal I}_{YMH}$ defined on the set of triple
$\{g_{\mu\nu},A_\mu,\varphi\}$. Here, $g_{\mu\nu}$ denotes the metric on
spacetime $\mm$ (and is usually assumed to be fixed by the flat metric),
$A_\mu$ denotes the gauge potential and $\varphi$ the
Higgs field. In contrast to the bosonic part of the standard
model, the fermionic action is defined with respect to a specific
Dirac operator $\di_{\!Y}$ called the "Dirac-Yukawa" operator. This
Dirac operator
takes the specific form\footnote{Our conventions are
specified in the next section.}:
\bb
\di_{\!Y} = i\gamma^\mu(\pa_\mu +\omega_\mu + A_\mu) + i\Phi,
\ee
where $\omega_\mu:=-\frac{1}{4}\,\gamma^{ab}\omega_{ab\mu}$ denotes
the spin connection defined by the Levi-Cevita connection on
spacetime $\mm$ (which drops out in the case of $g_{\mu\nu}$
being flat). $\Phi$ denotes an odd (anti-hermitian) matrix containing the
Higgs field $\varphi$ in a certain representation and defines
the "Yukawa coupling term":
\bb
i{\bar\psi}\Phi\psi \equiv
\sum_{a,b=1}^N{\bar\psi}_a(i\Phi)_{ab}\psi_b.
\ee
Here, the spin degrees of freedom of the fermions $\psi$ are
suppressed and N denotes the dimension of the fermion representation of
the gauge group $G$. Consequently, the fermionic action
${\cal I}_{DY}$  - the Dirac-Yukawa action -  can be considered as a
specific functional defined on the set
$\{g_{\mu\nu}, A_\mu,\varphi\}$ as well. This is the "usual" point of view
of how the action functional of the standard model is understood.
Of course, this point of view corresponds to the fact that the specific
form of the action serves as one of the main ingredients of a general
Yang-Mills-Higgs model building kit.

\smallskip

But there is also another perspective of how the action of the
standard model (without gravity!) can be viewed, namely the approach by
Connes and Lott which uses the framework of Connes' non-commutative geometry
(c.f. [C] and [CL]). In this approach the action functional is derived
from a "K-cycle". Moreover, Connes has also mentioned that the specific
Dirac operator:
\bb
\di = \gamma^\mu(\pa_\mu + \omega_\mu + A_\mu)
\ee
is linked to the euclidean Einstein-Hilbert action ${\cal I}_{EH}$ via
the Wodzicki residue of the inverse of $\di^2$. This was explicitly shown
in [K] and [KW]. Obviously, this Dirac operator defines the "kinetic" part
of the Dirac-Yukawa action. It might be worth mentioning that in the
Connes-Lott approach to the standard model the spin connection
$\omega_\mu$ always drops out and therefore the metric $g_{\mu\nu}$
on spacetime $\mm$ remains indefinite. If on the other hand, one uses
the kinetic part $\di$ of the Dirac-Yukawa operator (1) in
order to derive the Einstein-Hilbert action - as proposed by
Connes - the information contained in the gauge field $A_\mu$ is lost.

\smallskip

Following Connes' point of view, where a Dirac operator (K-cycle) is on
the input side and the action is on the output side, it is natural
to ask whether there is a Dirac operator $\dit$ from which both the
Einstein-Hilbert and the Yang-Mills-Higgs action can be derived
simultaneously. In this letter we answer this question affimatively.
Moreover, we show that this particular Dirac operator is completly
determined by the Dirac-Yukawa operator (1).

\smallskip

\section{Mathematical preliminaries}

In order to be specific and to fix our notation we
make the following assumptions: let spacetime $\mm$ be described by a
four dimensional compact Riemannian spin-manifold without
boundary. Then we use the following convention for the Clifford
relation: $\{\gamma^a,\gamma^b\}\equiv\gamma^a\gamma^b +
\gamma^b\gamma^a = -2g^{ab}$. The involution on the "spin-space"
$S$ ($\simeq\cc_L^2\op\cc_R^2$, locally) is denoted by
$\gamma_5$\ with the two conditions: $\gamma_5^2 = \eins_S$\ and
$\gamma_5^* = \gamma_5$. We also use the shorthand notation:
$2\gamma^{ab}:=[\gamma^a,\gamma^b]$. Furthermore it is
assumed that the "gamma matrices" $\gamma^a$\ define a hermitian
representation of the complexified Clifford algebra. Since we are
only interested in local statements we also assume that the (hermitian)
"gauge-bundle" $E\mapsto\mm$\ is trivial, i.e. we assume that
$E=\mm\times\cc^N$. In order to take parity violation into account
we set: $\cc^N=\cc^{n_1}\op\cc^{n_2}$ ($N:=n_1+n_2$). Consequently, the
internal space $E$\ is also $\zz_2-$graded and we denote the
corresponding involution on $E$\ by $\chi$.
Since the spinor fields $\psi=(\psi_L,\psi_R)$ are to be considered
as elements of the tensor product $\ep:=S\hot E$\ of the spin space
$S$ and the internal space $E\,$\footnote{I.e. $\psi = \psi_{sa}$, where
"s" denotes the spin degree of freedom and therefore refers to the
spin space $S$ and "a" denotes
the gauge degree of freedom and therefore refers to the
internal space $E$.} we can take into account the two $\zz_2-$gradings
of the respective spaces $S$ and $E$ by using the "graded tensor
product" - indicated by "$\hot$" - instead of the usual tensor
product\footnote{Let us recall that the
graded tensor product is defined as follows:
$(A\hot B)(A'\hot B'):= (-1)^{|B||A'|}AA'\hot BB'$, for all
$\,A,A',B,B'\in End(\ep),$ with $|B|,|A'|\in\{0,1\}$
depending on whether $B, A'$ are even or odd elements with respect to
$\chi$ and $\gamma_5$. The total grading on $\ep$ is defined by
$\gamma_5\hot\chi.$}. Thus $\psi(x)\in\ep_x =
S_x\hot E_x\simeq(\cc_L^2\op\cc_R^2)\hot\cc^N, x\in\mm$.
The unitary representation of the gauge group $G$ (which is
assumed to be  compact and semi-simple) on the internal space $E$ is
denoted by $\rho$, i.e.
$\rho:G\mapsto\hbox{\bf M}_N(\cc)\,$\footnote{From a mathematical
point of view most of these assumptions are redundant. For the general
setting we refer to [AT].}.

\smallskip

Since the covariant derivative\footnote{Here, ${\eins}_S$
and ${\eins}_E$ denotes the identity elements in
$End(S)\,(\simeq\hbox{\bf M}_4(\cc)$, locally) and
$End(E)\simeq\hbox{\bf M}_N(\cc)$, respectively.
Note that we have ignored the tensor product structure of $\ep$ in the
introduction.}
\bb
\nabla_\mu:= \pa_\mu + \omega_\mu\hot{\eins}_E + {\eins}_S\hot A_\mu
\ee
defining the Dirac operator $\di$ satisfies the crucial
relation:
\bb
[\nabla_\mu,\gamma^\nu] = -\gamma^\sigma\Gamma^\nu_{\sigma\mu},
\ee
it deserves a special name: it is called a Clifford connection. Here,
$\Gamma^\nu_{\sigma\mu}$\ denotes the Christoffel symbol. The
relation (5)  fixes the spin part of a Clifford connection to be
unambiguously defined by the Levi-Cevita connection. In other words:
two Clifford connections can differ only with respect to the gauge
field $A_\mu$.

\smallskip

As a consequence of (5) we can introduce the following covariant
section:
\bb
\xi_\mu:=-\frac{1}{4}g_{\mu\nu}\gamma^\nu\hot{\eins}_E
\ee
with the property of being covariantly constant with respect to any
Clifford connection (4). Additionally it fulfills:
\bb
\gamma^\mu\,\xi_\mu = {\eins}_\ep.
\ee

\smallskip

We now introduce the Wodzicki function ${\cal W}_\ep$. Let us denote
by ${\cal D}(\ep)$\ the affine space of all Dirac operators $\dit$
defined on $\ep$ which
satisfy\footnote{This relation can actually be considered as a
definition of a Dirac operator compatible with the given
Clifford structure on $\ep$\ since the space $\ep$\  not only denotes a
$\zz_2-$graded vectorbundle over $\mm$\ but also a "Clifford modul".}
\bb
[\dit,f] = \gamma^\mu\pa_\mu\,f
\ee
for all $f\in C^\infty(\mm)$.
Then the Wodzicki function is the particular functional on ${\cal D}(\ep)$
\bb
{\cal W}_\ep: {\cal D}(\ep)&\mapsto&\cc\cr
\dit&\mapsto&{\cal W}_\ep(\dit):= -\frac{1}{6N}Res(\dit^{-2}).
\eee
Here, $Res$\ denotes the Wodzicki residue which generalizes the
residue of Adler and Manin (c.f. [W], [A] and [M]). It can be shown
that $Res(\dit^{-2})$ is strongly related to the subleading term of the
asymptotic expansion of the heat kernel $exp(-\tau\dit^2)$\ of $\dit^2$
(c.f. [G] and [KW]). Moreover, using a generalized version of the
Lichnerowicz formula for the decomposition of $\dit^2$\ we have
shown that (c.f. [AT])
\bb
{\cal W}_\ep(\dit):=
\frac{1}{|\ep|}\int_\mm\!\!tr_\ep\left({\cal F}(\dit) -
\frac{r_\mm}{6}{\eins}_\ep\right)\sqrt{|g|}\,d^4x,
\ee
with $|\ep|:=\hbox{dim}\ep\,$ and $\,|g|:=\hbox{det}[g_{\mu\nu}]\,$.
Here, $r_\mm$ denotes the Ricci scalar of $\mm$ and the homomorphism
$\cal F$ takes the explicit form:
\bb
{\cal F}(\dit)\,&:=&\,\frac{1}{4}r_\mm\,{\eins}_\ep +
\frac{1}{2}\gamma^{\mu\nu}\hot F_{\mu\nu}\cr
&+& \frac{1}{2}[\gamma^\mu[\omega_\mu,\gamma^\nu],\omega_\nu] +
\gamma^{\mu\nu}('\nabla_\mu\omega_\nu) -
\frac{1}{2}\gamma^\mu[('\nabla_\nu\omega_\mu),\gamma^\nu]\cr
&+& \frac{1}{2}\gamma^{\mu\nu}[\omega_\mu,\omega_\nu] +
\frac{1}{4}g_{\mu\nu}\gamma^\sigma[\omega_\sigma,\gamma^\mu]
\gamma^\kappa[\omega_\kappa,\gamma^\nu],
\ee
with $\omega_\mu:=\xi_\mu(\dit-\di)$. Since
$\dit-\di\in End(\ep)$, here, the notation $\xi_\mu(\dit-\di)$ means
the product in $End(\ep)$. $\di$ is defined with
respect to any Clifford connection (4) and $F_{\mu\nu}$ denotes
the Yang-Mills field strength with respect to the gauge field
$A_\mu$.
Note that $A_\mu\in\rho_\ast(\cal G)\subset End(E)$, where
$\cal G$ denotes the Lie-algebra of the gauge group $G$ and
${\rho_\ast}$ the corresponding homomorphism between $\cal G$
and $End(E)$ induced by the representation $\rho$.
The covariant derivative $'\nabla_\mu$\ denotes the induced
Clifford connection on $T^*\mm\ot End(\ep)$, i.e.
\bb
'\nabla_\mu\,\omega_\nu: = [\nabla_\mu,\omega_\nu] -
\omega_\sigma\Gamma^\sigma_{\mu\nu}.
\ee
Note that the second and the third term in (10) always drop out when
calculating ${\cal W}_\ep$.

\smallskip

We are now in the position to define our approach to the
standard model. After we have fixed the general scheme  we shall introduce
a particular Dirac operator from which the full action functional of
the standard model can be derived.

\section{Model building kit}

\subsection{The general scheme}

In this subsection we introduce a new model building kit for gauge
theories similar to the Connes-Lott scheme without, however, using
the mathematical framework of non-commutative geometry
(c.f. [CL], [SZ], [KS] and [IS]). The main feature of the kit proposed
here is that it naturally includes gravity and therefore unifies the
latter with gauge theories.

\smallskip

The "input" of our model building kit is given by the following triple:
\bb
(G,\rho,\dit).
\ee
Again $G$ denotes a (compact, semi-simple) gauge group,
$\rho$ its representation concerning an internal space $E$ and $\dit$
any given Dirac operator on $\ep:=S\hot E$ satisfying the
relation (8), i.e. $\dit\in{\cal D}(\ep)$. The "rule" of our kit simply
consists in calculating the Wodzicki function ${\cal W}_\ep(\dit)$
concerning the Dirac operator $\dit$. This means that one has to
calculate the traces over the gamma matrices regarding the last four
terms in (10). The "output" is then a specific action functional for
the gauge theory in question. Of course, the chosen Dirac operator
also defines a particular fermionic action. This is our general scheme.
We now use this general building kit in order to derive the combined
Einstein-Hilbert-Yang-Mills-Higgs action. As we shall see this follows
from (10) in case that we usea specific Dirac operator $\dit$ which
also defines the Dirac-Yukawa action.

\subsection{The Einstein-Yang-Mills-Higgs building kit}

We define the following particular Dirac operator:
\bb
\dit:=
\pmatrix{\lambda_1\di_{\hat\Phi}&\gamma^{\mu\nu}{\hat F}_{\mu\nu}\cr
\phantom{0}&\phantom{0}\cr
-\gamma^{\mu\nu}{\hat F}_{\mu\nu}&\lambda_2\di_{\hat\Phi}}.
\ee
Here, the diagonal part of our Dirac operator is defined by the
Dirac-Yukawa operator $\di_{\hat\Phi}:=-i\di_{\!Y}$\ where we have
re-scaled the Yukawa coupling term
$\Phi\equiv{\eins}_S\hot\phi$ (with $\phi$ odd and anti-hermitian) by
$\hat\Phi:= a\Phi$.

\smallskip

The off-diagonal term is fixed by the element
\bb
{\hat F}_{\mu\nu}:=
[\nabla_\mu + \omega_\mu,\nabla_\nu + \omega_\nu] -
\frac{1}{4}R_{ab\mu\nu}\gamma^{ab}\hot{\eins}_E
\ee
which can be considered as a "relative curvature" on $\ep\,$\footnote{Actually,
it can be shown that this relative curvature is but the ``super curvature''
$F_{\phi}:=(\nabla^E + ib\phi)^2$ on the internal space $E$. Thus
the Higgs field can also be interpreted geometrically as a part of the
particular superconnection $\nabla^E + ib\phi$ on $E$ (c.f. [NS]). Note that
$\nabla^E + \phi$ uniquely defines the corresponding super connection of the
Dirac-Yukawa operator $\di_{\!Y}$.}.
Here, $\omega_\mu$\ is given by
\bb
\omega_\mu=-\frac{ib}{4}g_{\mu\nu}\gamma^\nu\hot\phi,
\ee
$R_{ab\mu\nu}$ denotes the Riemannian curvature on $\mm$ and
$\lambda_1,\lambda_2, a, b\,$ are real constants. Because of (15) the
Higgs field can be considered as defining a particular connection
${\tilde\nabla}_\mu = \nabla_\mu + \omega_\mu$\ on $\ep$. Consequently,
the Higgs field generalizes the Yang-Mills curvature on the gauge bundle
$E$, whereby the whole Dirac operator $\dit$ is fully defined with
respect to the Yukawa coupling term $\Phi$. Let us emphasizes that the
Dirac-Yukawa operator is also mathematically distinguished since it is
of "simple type" (c.f. [AT]). Therefore, the Dirac operator (13) is indeed
natural.

\smallskip

Regarding this particular Dirac operator (13) the Wodzicki function (9)
reads:
\bb
{\cal W}_{\ep}(\dit)\,&=&\,\alpha_o\lambda_1\lambda_2\,l_p^{-2}
\int_\mm\!\! r_\mm\,\sqrt{|g|}\,d^4x\cr &+&
\alpha_2\int_\mm\!\!tr(F^{\mu\nu}F_{\mu\nu})\,\sqrt{|g|}\,d^4x \cr &+&
\alpha_3 b^2
\int_\mm\!\!tr((\nabla_\mu\,\phi)^*\,\nabla^\mu\,\phi)\,\sqrt{|g|}\,d^4x\cr
&+&
\int_\mm\!\! tr(b^4\alpha_4(\phi^*\phi)^2 -
\alpha_1\lambda_1\lambda_2 a^2\, l_p^{-2}(\phi^*\phi))\,\sqrt{|g|}\,d^4x,
\ee
with $\alpha_o=1/12,\,\alpha_1=1/dimE,\,\alpha_2=(1/2)\alpha_1,
\,\alpha_3=(9/8)\alpha_1,\,\alpha_4=(3/2)\alpha_3$.

\smallskip

Here, the traces have to be performed within $End(E)$ and therefore depend
on the specific representation $\rho$\ which is chosen in the input.
Note that we have already introduced a physical length scale which is
fixed by the Planck length $\l_p$.

\smallskip

Of course, the fermionic part of the standard model is given by the
Dirac-Yukawa action:
\bb
{\cal I}_{DY} =
\int_\mm\!\!{\bar\psi}\,\di\psi\,\sqrt{|g|}\,d^4x +
b\int_\mm\!\!{\bar\psi}\,(i\Phi)\psi\,\sqrt{|g|}\,d^4x.
\ee
We remark that the relative constant in the off-diagonal term of the
Dirac operator $\dit$ is fixed by the relative sign between the
Einstein-Hilbert and the Yang-Mills action (which has to be
positiv here) and that the Dirac-Yukawa action must be real. Indeed,
the Dirac operator (13) is anti-selfadjoint.

\smallskip

Let us denote by ${\cal I}_{bosonic}$\ and by ${\cal I}_{fermionic}$
 the bosonic and the fermionic action, respectively. If we then define
$\Psi:=(\psi,\psi)$ we can summarize our result as follows:
\bb
{\cal I}_{bosonic} &:=& {\cal W}_\ep(\dit) \sim {\cal I}_{EHYMH}, \cr
{\cal I}_{fermionic} &:=& <\Psi,i\dit\Psi> \sim {\cal I}_{DY}.
\ee
These two action functionals will always be on the output side
when the Einstein-Yang-Mills-Higgs building kit is concerned:
\bb
(G,\rho,\dit),
\ee
with $\dit$ defined by (13). We emphasize that in this particular case the
Higgs representation is always contained in the fermionic representation.

\smallskip

\section{Conclusion}

We have shown that the whole action functional of the standard model and of
gravity can be derived from a certain Dirac operator. This
operator,in turn, is defined by the Yukawa coupling term generating the
masses of the fermions.

\smallskip

Moreover, our approach unifies gravity with Yang-Mills-Higgs gauge
theories. The Einstein-Hilbert action arises as a natural ``companion''
of the Yang-Mills-Higgs functional. Therefore, the metric in Yang-Mills
gauge theories is no longer fixed ``by hand'' but by the Einstein
equation, which is very satisfying, conceptually.

\smallskip

 From a geometrical point of view the full action functional of the
standard model can be considered as encoded within the subleading
term of the asymptotic expansion of the heat kernel corresponding to
the particular Dirac operator which we have introduced in the last
section. Consequently, it might be interesting also to investigate the
next terms of the expansion which could serve as a kind of correction
terms to the classical action.

\pagebreak

{\begref

\ref{[A]} M.Adler, Inent. Math. 50, 219-248 (1979) {\sl On a trace
functional for formal pseudo-diff. operators and the symplectic
structure of Korteweg-de Vries type equations}.

\smallskip

\ref{[AT]} T.Ackermann, J.Tolksdorf,  \sl
The generalized Lichnerowicz formula and analysis of Dirac operators
\rm ,
Preprint CPT-94/P.3106 and Mannheimer Manuskripte 181, (1994)

\smallskip

\ref{[C]} A.Connes, \sl Non-commutative geometry and physics\rm ,
to be published in the proceedings of the 1992 Les Houches Summer
School.

\smallskip

\ref{[CL]} A.Connes, J.Lott,\sl Particle models and non-commutative
geometry,\rm Nucl. Phys. B Proc. Supp. {\bf 18B} (1990) 29-47.

\smallskip

\ref{[G]} P.Gilkey, {\sl Invariance Theory, The Heat Equation, And the
Atiyah-Singer Index Theorem}, Publesh or Perish (1984).

\smallskip

\ref{[IS]} B.Iochum, Th. Sch\"ucker, {\sl Yang-Mills-Higgs versus
Connes-Lott}, CPT-94/P.3090.

\smallskip

\ref{[K]} D.Kastler, \sl The Dirac operator and gravitation\rm ,
to appear in Com. Math. Phys.

\smallskip

\ref{[KS]} D.Kastler, Th.Sch\"ucker, {\sl A detailed account of Alain
Connes' version of the standard model in non-commutative geometry},
4,CPT-94/P.3092 (1994).

\smallskip

\ref{[KW]} W.Kalau, M.Walze, \sl Gravity, non-commutative geometry
and the Wodzicki residue\rm , to appear in Journ. of
Geometry and Physics.

\smallskip

\ref{[M]} Yu.I.Manin, J. Soviet Mathematics 11, 1-122 (1979),
{\sl Algebraic aspects of non-linear differential equations}.

\smallskip

\ref{[NS]} Y.Ne'eman, S.Sternberg,{\sl Internal Supersymmetry and
Superconnections}, Symplectic Geometry and Math. Physics, Ed. by
P.Donato, C.Durce, J.Elhadad, G.M.Tuymmann, Birkh{\"a}user 1991.

\smallskip

\ref{[SZ]} Th.Sch\"ucker, J.-M.Zylinski, {\sl Connes' model building kit},
J. Geom. Phys. 16 (1994) 1.

\smallskip

\ref{[W]} M.Wodzicki, \sl Non-commutative residue I\rm , LNM {\bf 1289}
(1987), 320-399.}

\end{document}